\documentclass{aa}\usepackage{graphics}
\begin{document}
\thesaurus{06(06.13.1; 02.03.1; 02.13.1)}
\title{The influence of geometry and topology on axisymmetric mean-field
dynamos}
\author{Eurico Covas\thanks{e-mail: eoc@maths.qmw.ac.uk}\inst{1}
\and Reza Tavakol\thanks{e-mail: reza@maths.qmw.ac.uk}\inst{1}
\and Andrew Tworkowski\thanks{e-mail: A.S.Tworkowski@qmw.ac.uk}\inst{2}
\and Axel Brandenburg\thanks{e-mail: Axel.Brandenburg@ncl.ac.uk}\inst{3}
\and John Brooke\thanks{e-mail:zzalsjb@afs.mcc.ac.uk}\inst{4}
\and David Moss\thanks{e-mail:moss@ma.man.ac.uk}\inst{5}}
\institute{Astronomy Unit, School of Mathematical Sciences,
Queen Mary and Westfield College, Mile End Road, London E1 4NS, UK
\and Mathematical Research Centre, School of Mathematical Sciences,
Queen Mary and Westfield College, Mile End Road, London E1 4NS, UK
\and Department of Mathematics, University of Newcastle
upon Tyne NE1 7RU, UK
\and Manchester Computing, The University, Manchester M13 9PL, UK
\and Department of Mathematics, The University, Manchester M13 9PL, UK}
\date{Received ~~ ; accepted ~~ }
\offprints{\it http://www.maths.qmw.ac.uk/$\sim$eoc}
\maketitle
\markboth
{Covas {\em et al.}: Mean-field dynamos with different geometries
and topologies}
{Covas {\em et al.}: Mean-field dynamos with different geometries
and topologies}
\begin{abstract}
We study the changes in the dynamical behaviour of axisymmetric
spherical mean-field dynamo models produced by changes in their
geometry and topology, by considering a two parameter family of models,
ranging from a full sphere to spherical shell, torus and disc--like
configurations, within a unified framework.
We find that the two parameter space of the family of models
considered here separates into at least three different regions with
distinct characteristics for the onset of dynamo action. In two of
these regions, the most easily excited fields are oscillatory, in one
case with dipolar symmetry, and in the other with quadrupolar, whereas in
the third region the most easily excited field is steady and
quadrupolar.
In the nonlinear regime, we
find that topological changes can alter significantly the dynamical
behaviour, whilst modest changes in geometry can produce qualitative
changes, particularly for thin disc--like configurations.  This is of
potential importance, since the exact shapes of astrophysical bodies,
especially accretion discs and galaxies, are usually not precisely known.
\end{abstract}
\section{Introduction}
\label{intro}
The magnetic fields observed in stars, accretion discs and galaxies are
generally considered to be generated by magnetohydrodynamic dynamo
action.  The complexity of the physics of such dynamos has meant that a
fully self-consistent model is beyond the range of the computational
resources currently available, although important attempts have been
made to understand turbulent dynamos in stars (Nordlund {\em et al.} 1992,
Brandenburg {\em et al.} 1996) and accretion discs (Brandenburg {\em et al.}
1995, Hawley {\em et al.} 1996), for example.  Such studies have had to be
restricted to the geometry of a Cartesian box, thus they are in essence
local dynamos. However, magnetic fields in astrophysical objects are
observed to exhibit large scale structure, related to the shape of the
object, and thus can only be captured fully by global dynamo models.
In some cases it has been possible to show that,
for fully three-dimensional turbulence simulations,
the dependence of the results (parity, time dependence, cycle period, etc.)
on global
properties including boundary conditions is remarkably similar to that of
the corresponding mean-field dynamo models (Brandenburg 1999).
This lends some support to
the validity of mean-field theory, which is used in the following.

Mean field dynamo models, which attempt to capture such geometric
features, have been employed extensively in order to study various
aspects of the dynamics of solar, stellar and galactic dynamos (see for
example, Steenbeck \& Krause 1969, Roberts \& Stix 1972).  Often the
dimensionality of the underlying partial differential equations has
been reduced by assuming axisymmetry.  As the presence of strong
differential rotation tends to destroy non-axisymmetric fields (e.g.\
R\"adler 1986), this restriction has some astrophysical justification,
but nevertheless some galaxies and types of stars are known to have
non-axisymmetric fields and so this simplification cannot be
universally valid.

In addition to producing solutions with pure and mixed parities
(Brandenburg {\em et al.} 1989a,b), these studies have also revealed new
features such as solutions with both chaotic behaviour (see for example
Jones {\em et al.} 1985 in a severely truncated system, Brooke \& Moss 1994,
Torkelsson \& Brandenburg 1994b,
Tavakol {\em et al.} 1995) and intermittent modes of behaviour (see e.g.
Brooke \& Moss 1995; Tobias \& Weiss, 1997; Tworkowski {\em et al.} 1998,
Covas {\em et al.} 1998a).

An important shortcoming of these models is that they involve severe
approximations, which essentially fall into two broad categories. There
are those concerning the underlying physics, such as approximations
involved in the parameterisation of the turbulent processes in solar
and stellar convective zones (which are not known precisely), and there
are the simplifications assumed in modelling the phenomenological
features of these systems, such as the details of their geometries.
These shortcomings are often ignored, the assumption being that
changing the details of these models should not change their behaviours
qualitatively. A priori, however, there are no reasons why this should
be so and it is therefore important that this assumption should be
checked.

Recently, an attempt was made to study the robustness of axisymmetric
mean-field models in spheres and spherical shells with respect to
changes in the functional form of the $\alpha$ effect (Tavakol {\em et
al.} 1995) and also by considering the $\alpha$--quenching to be dynamic
(Covas {\em et al.} 1998a).  Here we shall employ the usual algebraic
$\alpha$--quenching and focus on the second feature discussed above,
namely the assumptions regarding the topology and geometry of these
models.  This is important for two reasons. Firstly, these are the two
important features differentiating the astrophysical situations in which
such dynamos are thought to be operative.  Thus spherical and spherical
shell models are considered in connection with solar and stellar
variability (with the sphere or shell representing convectively
unstable regions of stars), whereas disc (Torkelsson \& Brandenburg
1994a,b, 1995) and torus (Brooke \& Moss 1994; 1995) models are
considered in connection with modelling accretion discs and galaxies.
Secondly, the geometries of these models are often only approximations
to the true shapes of the astrophysical systems, and so it is important
to know the sensitivity of the dynamo solutions to modest changes in
geometry. This is particularly true for dynamo models in accretion discs
and galaxies.

In this paper we study the qualitative nature of the dynamical behaviour
of axisymmetric mean-field dynamo models as a function of changes in
their topology, following transitions from sphere to shell to torus, and
as a function of changes in their geometry resulting from alteration of
the shapes of the dynamo region, whilst retaining the topology.  We
consider these changes within a unified framework, using the same
numerical code with the same boundary conditions, to solve the same
dynamo equations. In the context of disc-like models, we also consider
changes in the rotation law.

The models considered here include geometric features such as
curvature, which are naturally present in astrophysical systems. There
have also been studies of more idealized models in Cartesian geometry,
and of low order models, which aim to describe generic features of
axisymmetric dynamo solutions (e.g.\ Weiss, Cattaneo \& Jones 1984,
Jennings {\em et al.} 1990, Jennings \& Weiss 1991, Jennings 1991,
Tobias {\em et al.} 1995).  It is of interest to determine whether such
features persist across changes in geometry and topology and in the
highly nonlinear regimes considered here.  We shall briefly return to
this issue below in the light of our findings.

In section 2 we briefly introduce the equations for the axisymmetric
mean-field dynamo models together with the relevant boundary
conditions.  Section 3 contains a discussion of our results and finally
section 4 contains our conclusions.
\section{The model}
The standard mean-field dynamo equation (cf.\ Krause \& R\"adler 1980) is
of the form
\begin{equation}
\label{dynamo}
\frac{\partial \vec{B}}{\partial t}=\nabla \times \left( \vec{u}
\times \vec{B} + \alpha \vec{B} - \eta_t \nabla \times \vec {B} \right),
\end{equation}
where $\vec{B}$ and $\vec{u}$ are the mean magnetic field and the mean
velocity respectively. The quantities $\alpha$ (the $\alpha$ effect)
and the turbulent magnetic diffusivity, $\eta_t$, appear in the process
of the parameterisation of the second order correlations between the
velocity and magnetic field fluctuations ($\vec{u}'$ and $\vec{B}'$).
For the sake of simplicity, and to facilitate comparison with previous
work, we shall ignore anisotropies and take $\alpha$ and $\eta_t$ to be
scalars.  We assume that ${\bf u}={\bf \Omega}\times{\bf
r}=u_\phi\hat\phi$.  Nondimensionalization of the equations in terms of a
length $R$ and a time $R^2/\eta_t$ produces the two usual dynamo
parameters $C_\alpha=\alpha_0 R/\eta_t$ and $C_\omega=\Omega_0
R^2/\eta_t$, where $\alpha_0$ and $\Omega_0$ are typical values of
$\alpha$ and $|{\bf \Omega}|$.  For the sake of comparison with existing
results (e.g., Covas {\em et al.} 1998a),
we set $|C_{\omega}|=10^4$ throughout (except briefly, in one particular case).
We use spherical polar
coordinates $(r,\theta,\phi)$ and consider axisymmetric solutions only.
In order to satisfy the condition $\nabla\cdot{\bf B}=0$ we solve Eq.\
(\ref{dynamo}) by splitting the field into meridional and azimuthal
components, ${\bf B} = {\bf B}_{p} + {\bf B}_{\phi}$, and expressing
these components in terms of scalar field functions, so that ${\bf
B}_{p} = \nabla \times a \hat{\mbox{\boldmath $\phi$}}$, ${\bf B}_{\phi}
= b \hat{\mbox{\boldmath $\phi$}}$.

We consider the usual algebraic form of $\alpha$--quenching namely
\begin{equation}\label{alpha_a}
\alpha_a=\frac{\alpha_0 \cos \theta}{1+ \vec{B}^2},
\end{equation}
with $\alpha_0={\rm constant}$. This form has been adopted in numerous
studies.

In our calculations we consider two types of rotation laws, one
with constant shear and the other with approximately rigid rotation
close to the rotation axis and tending to Keplerian rotation at larger
radii. Hereafter we refer to the latter as a ``generalized Keplerian''
rotation law.  The first constant shear profile has been used
extensively in studies in many different geometries. We recognise that
helioseismological evidence indicates that the rotational profile of
the sun gives a shear in a very concentrated region. Our main aim is to
allow comparison with a range of published models, hence we chose this
simple and widely used form. The second profile has been extensively
used in studies of accretion discs. In that it gives a
differential rotation that decreases with radius it might be considered
to give a generic representation of the rotation in a galactic disc,
although rotation profiles in galaxies are certainly not Keplerian.

For the rotation profiles that we will consider here, we take
\begin{equation}\label{constant}
\vec{u}=\Omega \varpi \hat{\mbox{\boldmath $\phi$}},
\end{equation}
where $\varpi=r\sin\theta$ is the distance from the rotation axis.
For the first profile mentioned above,
we shall, consistent with Brandenburg {\em et al.} (1989a),
take $\Omega=\Omega_0 r$, where $\Omega_0$ is constant,
and for the second profile,  we shall take
\begin{equation}
\Omega = \Omega_0 \left [ 1 + \left ( \frac {\varpi}
{\varpi_0} \right )^{qn}\right ]
 ^{-1/n},
\label{kepler}
\end{equation}
where $q=3/2$ for accretion
discs. In conformity with Torkelsson \& Brandenburg
(1994a), we take $\varpi_0$ to be $0.3$ and $n=10$. Relation
(\ref{kepler}) approximates solid body rotation for small $\varpi$ and
approaches Keplerian rotation for $\varpi\gg\varpi_0$. Note that for
this profile $\partial\Omega/\partial r<0$ when $C_\omega>0$, whilst for
profile (\ref{constant}) with constant shear the signs are reversed, i.e.\
$\partial\Omega/\partial r>0$ when $C_\omega>0$.

\begin{figure}[!thb]
\resizebox{\hsize}{!}{\includegraphics{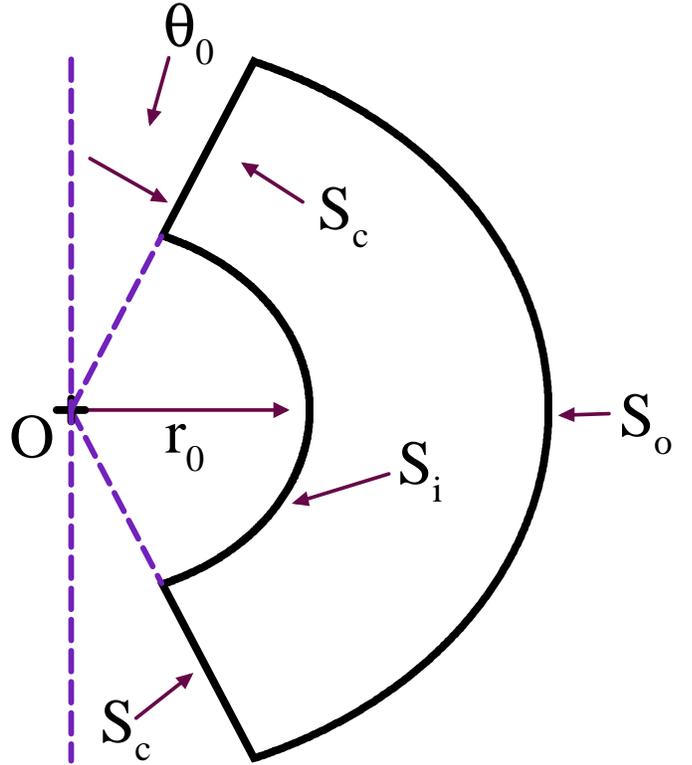}}
\caption{\label{boundary}
A schematic representation of the dynamo region
produced by the cuts characterised by the
model parameters $r_0$ and $\theta_0$.}
\end{figure}

The models we shall consider here are all constructed from a complete
sphere of radius $R$. We can remove an inner concentric sphere of
radius $r_0$ (in order to go smoothly from a sphere to a spherical shell). The
change from a sphere or a spherical shell to a torus- or disc-like
configuration is produced by excising a cone of semi-angle $\theta_0$
about the rotation axis, from the north and south polar regions. The
configuration is illustrated in Fig.\ \ref{boundary},
where the dynamo region is produced by revolving the meridian section shown
about the vertical dotted rotation axis through the origin $O$.

In order to solve the dynamo equations, boundary conditions need to be
imposed on the surfaces of the dynamo region created by removing the
cones (denoted by $S_{c}$), on the inner surface $r=r_0$ (denoted by
$S_{i}$), and also on the outer curved boundary (denoted by $S_{o}$), as
depicted in Fig.\ \ref{boundary}.

We adopt a perfect conductor boundary condition,
\begin{equation}\label{boundary2}
a=0, \quad
{{\partial (rb)}\over {\partial r}}=\alpha
{{\partial (ra)}\over {\partial r}}\quad\mbox{on}\quad S_{i}.
\end{equation}
Other choices are possible (see, e.g., Tworkowski {\em et al.} 1998), but
this has been used commonly in earlier investigations.

On $S_o$ a vacuum boundary condition has often been used. However this,
although appealing, is actually quite arbitrary as, for example, there
is no evidence that the solar poloidal field is curl-free immediately
above the surface. When $\theta_0>0$, implementing an essentially
non-local vacuum boundary condition is not straightforward, and so we
have chosen a plausible local condition.  Other authors have previously
taken a radial field condition, $\partial(ar)/\partial r = 0$, (e.g.\
Gilman \& Miller 1981). We used
\begin{equation}
\partial a/\partial r=0, b=0,
\end{equation}
which results in a poloidal field on $S_o$ which is typically near to
radial over most of the boundary. We stress that, in the absence of a
detailed model of processes above a stellar surface, all choices of
boundary condition on $S_o$ are more-or-less arbitrary, within general
restrictions of plausibility.

If a perfect conductor boundary condition is used on the surfaces
$S_{c}$ then as $\theta_0\rightarrow0$ the model does not tend to that
used for the full sphere, where $a=b=0$ on the axis. Thus we set
\begin{equation}
a=b=0\quad\mbox{on}\quad S_{c},
\end{equation}
to ensure continuity of behaviour as $\theta_0 \rightarrow 0$; again
other choices are possible.

In the following, as is customary, we shall discuss the behaviour of
the solutions by monitoring the total magnetic energy,
$E={1\over2\mu_0}\int\vec{B}^2dV$, taken over the dynamo region given
by $r_0 \le r \le R$ and $\theta_0\le \theta \le \pi - \theta_0$. We
split $E$ into two parts, $E=E^{(A)}+E^{(S)}$, where $E^{(A)}$ and
$E^{(S)}$ are respectively the energies of those parts of the field
whose toroidal field is antisymmetric and symmetric about the equator.
The overall parity $P$ is given by $P=[E^{(S)}-E^{(A)}]/E$ (Brandenburg
{\em et al.} 1989a), so $P=-1$ denotes an antisymmetric (dipole-like) pure
parity solution and $P=+1$ a symmetric (quadrupole-like) pure parity
solution.

For the numerical results reported in the following sections, we used a
modified version of the axisymmetric dynamo code of Brandenburg {\em et
al.} (1989a).  We took a grid size of $41\,\times\,81$ mesh
points in the dynamo region.  To test the robustness of the code we
verified that no qualitative changes were produced by employing a finer
grid, different temporal step length (we used a step length of
$10^{-4}R^2/\eta_t$ in the results presented in this paper).  We
considered a family of models ranging from a full sphere through
spherical shells to torus- and disc-like configurations.

\section{ Results}
We present the results of our comparative studies of dynamo solutions
in regions that are produced by variations of $r_0$ and $\theta_0$ in
the ranges $0\le r_0 < 1$ (in dimensionless units)
and $0\le \theta_0 < \pi/2$ respectively.
In particular, we took the following values:
\begin{eqnarray}
r_0&=& 0,~0.2,~0.5,~0.7\\
\theta_0&=&0,~4.5^{\circ},~22.5^{\circ},~45^{\circ},~75^{\circ}
\end{eqnarray}
To present our results, we use the following notation in the figures:
the prefix ``steady'' denotes a constant energy, whilst ``A'' and ``S''
on their own respectively represent pure antisymmetric (dipolar) ($P=-1$)
and
symmetric (quadrupolar) ($P=+1$) solutions with periodically
oscillating energy, ``OM'' represents solutions which possess periodic
oscillations in both parity and energy, and ``C'' and ``I'' denote
chaotic and intermittent behaviours respectively.
By intermittent behaviour we loosely mean irregular
switchings between two or more different dynamical regimes characterised
by different statistics. (For a precise characterisation of this type of
behaviour in some dynamo models see Covas {\em et al.} 1998a.)
We should add here that the computational cost of our calculations
limited the resolution to which we were able to search
the parameter space.
Consequently all our results regarding the sequence of
parity changes are subject to this limitation.

Since many previous authors have discussed behaviour at the onset of
dynamo action, we shall present our results in two subsections.  First
we present our results concerning the onset of dynamo action as we go
from a sphere through spherical shells to torus- and disc-like
structures, and then we consider the dynamical behaviour for each
configuration as the dynamo number is increased.  Our detailed results
are presented in Table \ref{onsetenergy} and Figs.\ \ref{onset} to \ref{transition.planar}.
\subsection{The onset of dynamo action}
\begin{figure}[!thb]
\resizebox{\hsize}{!}{\includegraphics{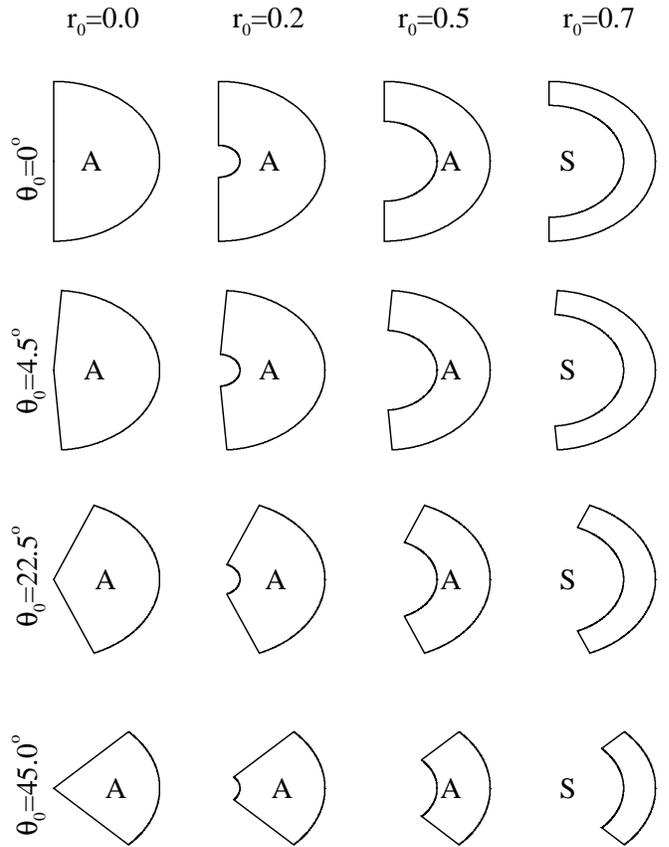}}
\caption{\label{onset}
Schematic representation of the dynamo regions and
their corresponding onset parity with $C_{\omega} = -10^4$ and
a constant rotational shear profile.}
\end{figure}

We summarise our onset results by considering the two types of
deformation of the basic sphere in turn.  Starting with the case where
an inner sphere is removed, giving a spherical shell, we note that
Roberts (1972) found numerically (but for different boundary
conditions) that for an $\alpha \omega$ dynamo with a negative dynamo
number, the most easily excited mode has dipole-like
symmetry for $r_0 <0.7$, whereas for $r_0\ge 0.7$ the most easily excited mode has
quadrupole-like symmetry.  Our corresponding results are summarised in
Fig.\ \ref{onset}. In particular, the top row of this figure,
summarizing the behaviour of a sphere and spherical shells without polar
cones removed, agrees qualitatively with the results of Roberts (1972).

\begin{figure}[!thb]
\resizebox{\hsize}{!}{\includegraphics{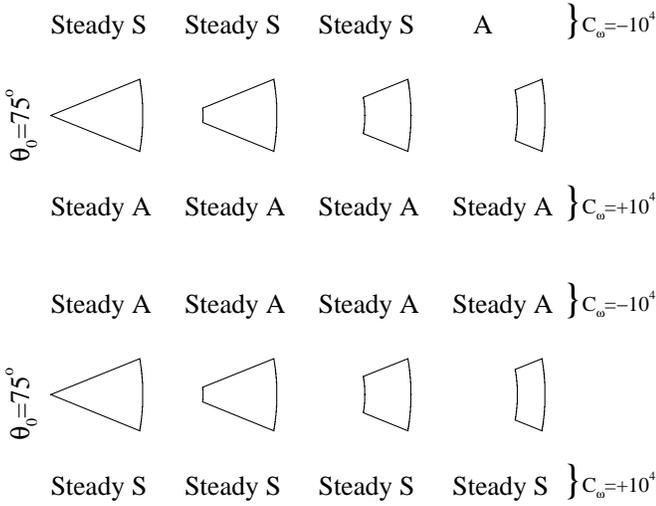}}
\caption{\label{onset_large}
Schematic representation of the dynamo regions and
their corresponding onset parity. The top and bottom panels are for
constant shear and  generalized Keplerian rotational profiles
respectively.}
\end{figure}

The next three rows in this figure correspond to the removal of cones
of semi-angles $\theta_0 = 4.5^{\circ},22.5^{\circ}$ and $45^{\circ}$
respectively. These results indicate that despite the removal of the
polar cones, the onset behaviour still remains the same as that given in
Roberts (1972).  In this way we may say that the onset results for
spheres are quite robust with respect to the polar cuts considered
here.

\begin{table}
\caption{\label{onsetenergy}
Values of $C_{\alpha}$ for the onset of dynamo action as a function of
the shell radius $r_0$ and the cut angle $\theta_0$ with $C_{\omega}
=-10^4$. The row labeled $*$ is produced using the generalized Keplerian
rotation profile given by Eq.\ (\ref{kepler}) with
$C_\omega = -10^4$. Here $SS$ indicates a
Steady S state.}
\begin{flushleft}
\begin{tabular}{lcccc}
\hline\noalign{\smallskip}
$\theta$ & $r_0=0.0$ & $r_0=0.2$ & $r_0=0.5$ & $r_0=0.7$ \\
\noalign{\smallskip}
\hline\noalign{\smallskip}
$0$           & 0.45  A & 0.38   A & 0.31  A & 0.42  S \\
$4.5^{\circ}$ & 0.45  A & 0.38   A & 0.32  A & 0.42  S \\
$22.5^{\circ}$& 0.70  A & 0.60   A & 0.42  A & 0.50  S \\
$45^{\circ}$  & 1.7   A & 1.3    A & 0.9   A & 1.0   S \\
$48^{\circ}$  & 1.86  A & 1.48   A & 1.04  A & 1.07  S \\
\hline\noalign{\smallskip}
$49^{\circ}$  & 1.98  A & 1.58   A & 1.10  A & 1.113 A \\
$50^{\circ}$  & 2.1   A & 1.679  A & 1.15  A & 1.17  A \\
$55^{\circ}$  & 3.0   A & 2.37   A & 1.56  A & 1.493 A \\
$60^{\circ}$  & 4.45  A & 3.545  A & 2.27  A & 1.98  A \\
$65^{\circ}$  & 7.2   A & 5.72   A & 3.57  A & 2.77  A \\
$67^{\circ}$  & 9.5   A & 7.1    A & 4.4   A & 3.24  A \\
\hline\noalign{\smallskip}
$70^{\circ}$  & 11   SS & 8.05  SS & 3.09 SS & 4.295 A \\
$75^{\circ}$  & 13.0 SS & 10.1  SS & 5.8  SS & 7.6   A \\
\hline\noalign{\smallskip}
$75^{\circ}*$ & 4.0  SA & 3.7   SA & 3.5  SA & 11.5 SA \\
\hline\noalign{\smallskip}
\end{tabular}
\end{flushleft}
\end{table}

To compare our results further with those from in previous studies,
for example of
dynamo action in tori by Brooke \& Moss (1994, 1995) and in
discs by Torkelsson \& Brandenburg (1994a,b, 1995), we also studied the
case of a large polar cut angle ($\theta_0 = 75^{\circ}$),
with both constant shear
and generalized Keplerian rotation profiles, and these results are
summarized in Fig.\ \ref{onset_large}.  Note that, as commented after
Eq.\ (\ref{kepler}), with the generalized Keplerian rotation law (\ref{kepler}),
$C_\omega>0$ corresponds loosely to the constant shear law with
$C_\omega<0$.  However equal values of $|C_\omega|$ do not imply equal
values of the rotational shear.

As can be seen from a comparison of Figs.\ \ref{onset} and
\ref{onset_large}, the onset behaviour in these models changes from
oscillatory to steady solutions as the cut angle $\theta_0$ increases.
Such a change in behaviour has also been found in oblate spheroidal
geometry as a sphere flattens to a disc-like configuration (Stix
1975, Soward 1992a,b and Walker \& Barenghi 1994).  According to these
authors, the crucial factor in the change from oscillatory to steady
behaviour is the aspect ratio of the spheroids, defined as the ratio
of the minor to the major axis. They speculated
that this was because dynamo waves were restricted from propagating in
the vertical direction (i.e.\ parallel to the minor axis).  In this
context it is interesting to note that in our case the only oscillating
solution for $\theta_{0}$ = $75^{\circ}$ occurs when $r_{0} = 0.7$ and
the above aspect ratio is $O(1)$, as for a sphere.

In Table~\ref{onsetenergy} we summarise the values of $C_{\alpha}$
corresponding to the onset of dynamo action in each configuration.
Except for the case with generalized Keplerian rotation law, this
shows that for a fixed $r_0$ the values of $C_{\alpha}$ at the onset
of dynamo action increase
as the cut angle $\theta_0$ increases for almost all configurations as
the shells get thicker.

We should also note that the steep increase in the onset value of
$C_{\alpha}$, as we go from $\theta_0 =45^\circ$ to $\theta_0 =
75^\circ$, may be related to the transition from shell--like to
disc--like configurations.  Table~\ref{onsetenergy} in conjunction with
Figs.\ \ref{onset} and \ref{onset_large} confirm that in the constant
shear case configurations with smaller polar cut angles follow the
linear results due to Roberts (1972), while those with larger cut angles
($\theta_0 \ga 49^{\circ}$) do not, thus
suggesting the possible existence of a transition as the cut angle is
increased.

For the disc--like configurations (with $\theta_0 \ga 70^{\circ}$), the
onset of dynamo action is to a steady state when $r_0<0.7$, when the
symmetry is determined by the sign of $\frac{\partial \Omega}{\partial
r}$.  If this is positive, that is for the constant shear with
$C_{\omega} = +10^4$ and for the generalized Keplerian rotation profile
with $C_{\omega} = -10^4$, the onset modes are antisymmetric.  For the
negative case the onset mode is always symmetric.

\subsection{Nonlinear regime}
\begin{figure}[!thb]
\resizebox{\hsize}{!}{\includegraphics{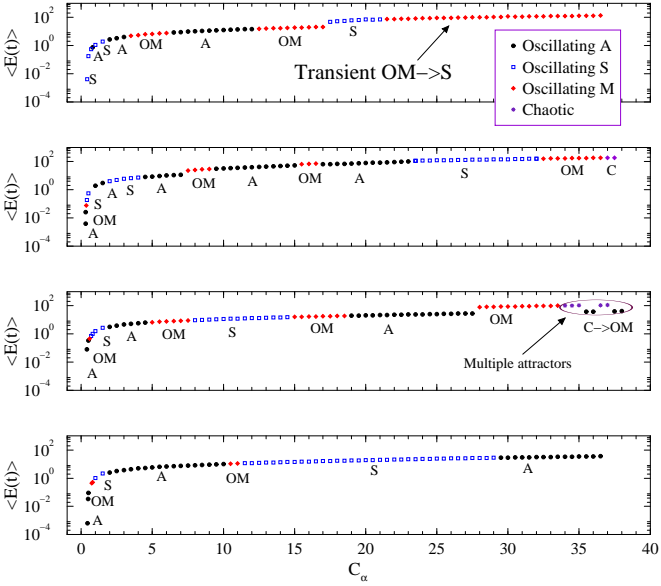}}
\caption{\label{none}
Partial bifurcation diagram, showing the energy in the stable solution, for
the case of complete spherical shells ($\theta_0=0$), with constant
(negative) shear, $C_{\omega}=-10^4$. For oscillatory modes, the mean
energy is plotted. Here and in following figures, the panels from top to
bottom correspond to $r_0$ values $0.7, 0.5, 0.2$ and $0.0$
respectively.}
\end{figure}

\begin{figure}[!thb]
\resizebox{\hsize}{!}{\includegraphics{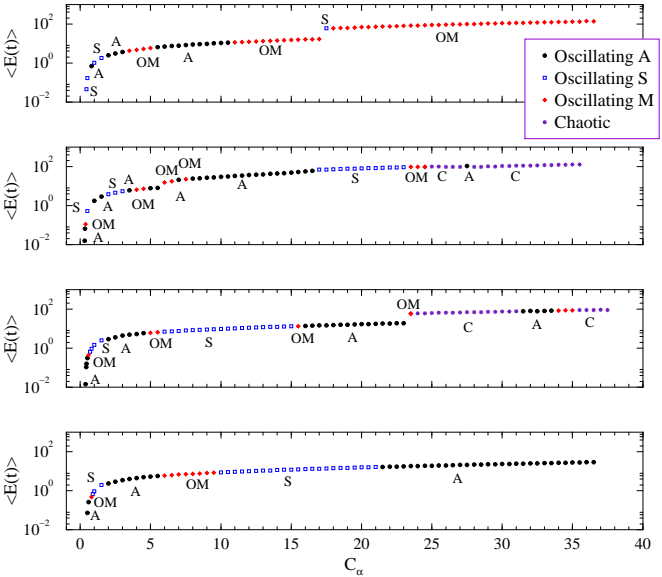}}
\caption{\label{very_small}
Partial bifurcation  diagram, as Fig.\ \ref{none}, for the case of a polar cut
with a very small angle
$\theta_0=4.5^{\circ}$ and
with constant (negative) shear, $C_{\omega}=-10^4$.}
\end{figure}

\begin{figure}[!thb]
\resizebox{\hsize}{!}{\includegraphics{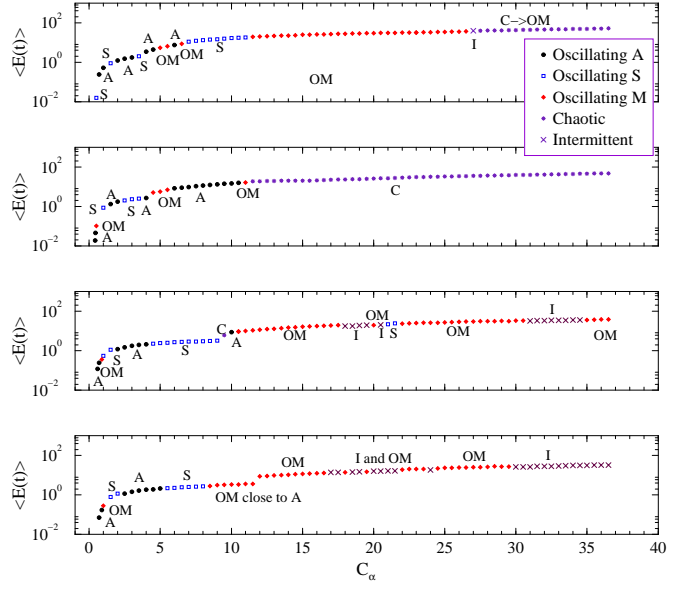}}
\caption{\label{small}
Partial bifurcation  diagram, as Fig.\ \ref{none}, for the case
$\theta_0=22.5^{\circ}$
with constant (negative) shear, $C_{\omega}=-10^4$.}
\end{figure}

\begin{figure}[!thb]
\resizebox{\hsize}{!}{\includegraphics{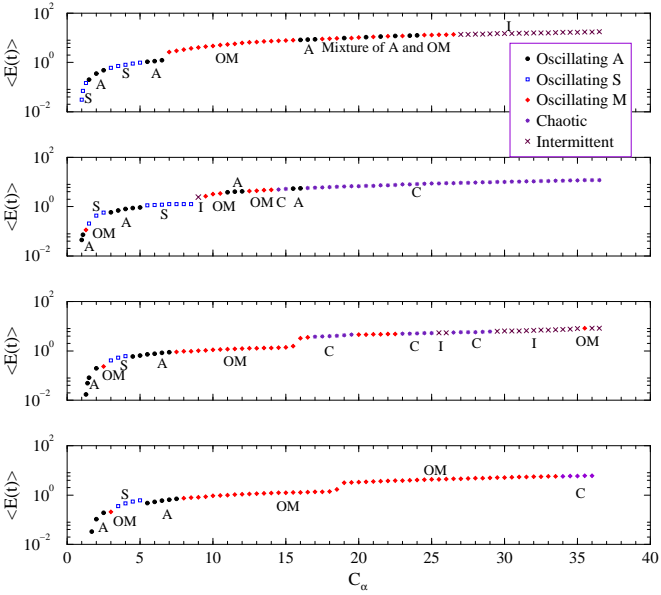}}
\caption{\label{medium}
Partial bifurcation  diagram, as Fig.\ \ref{none}, for the case
$\theta_0=45^{\circ}$,
with constant (negative) shear, $C_{\omega}=-10^4$.}
\end{figure}

\begin{figure}[!thb]
\resizebox{\hsize}{!}{\includegraphics{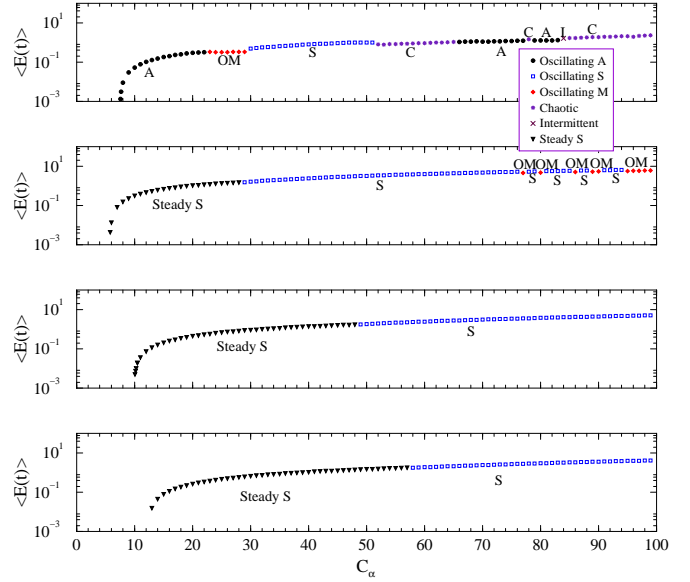}}
\caption{\label{largeminus}
Partial bifurcation  diagram, as Fig.\ \ref{none}, for the case of a large polar
cut angle, $\theta_0=75^{\circ}$,
with constant (negative) shear, $C_{\omega}=-10^4$.}
\end{figure}

\begin{figure}[!thb]
\resizebox{\hsize}{!}{\includegraphics{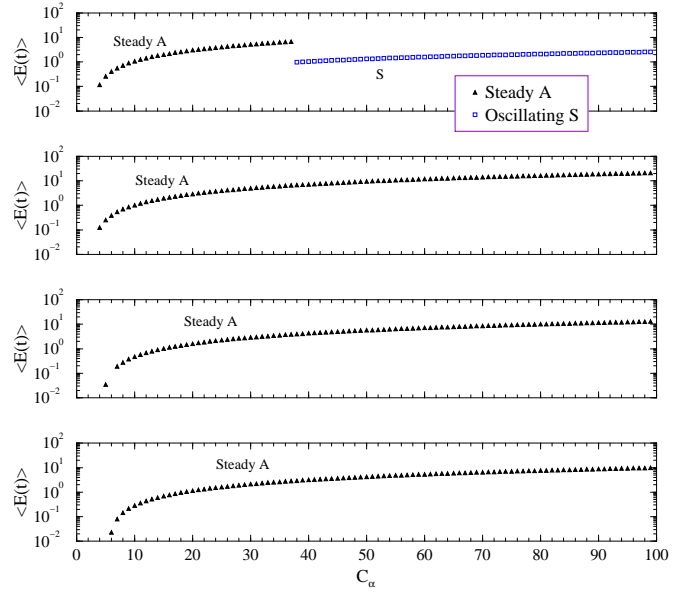}}
\caption{\label{largeplus}
Partial bifurcation  diagram, as Fig.\ \ref{none}, for the case of a large polar
cut angle, $\theta_0=75^{\circ}$,
with constant (positive) shear, $C_{\omega}=+10^4$.
}
\end{figure}

\begin{figure}[!thb]
\resizebox{\hsize}{!}{\includegraphics{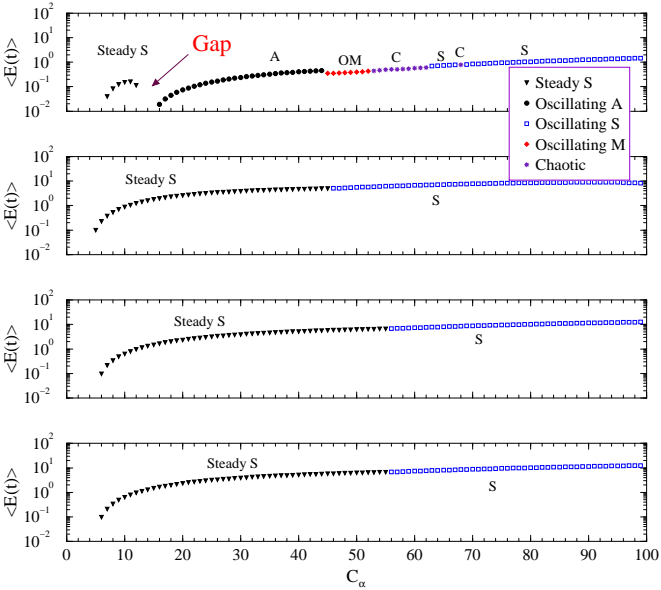}}
\caption{\label{keplerianplus}
Partial bifurcation  diagram, as Fig.\ \ref{none}, for the case of a large polar
cut angle, $\theta_0=75^{\circ}$,
with the generalized Keplerian rotation law, $C_{\omega}=+10^4$
($\partial\Omega/\partial r<0$),
$\varpi_0=0.3$.}
\end{figure}

\begin{figure}[!thb]
\resizebox{\hsize}{!}{\includegraphics{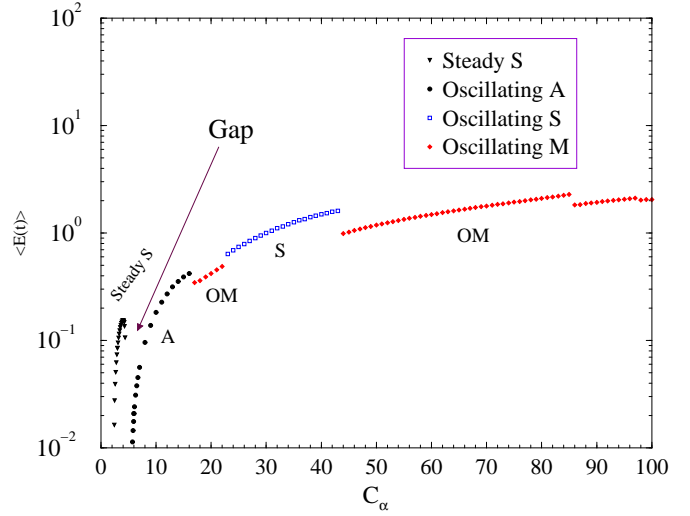}}
\caption{\label{keplerianplus2}
Partial bifurcation diagram, as Fig.\ \ref{none}, for the case of a large polar
cut angle, $\theta_0=75^{\circ}$,
with the generalized Keplerian rotation law, $C_{\omega}=+2.7\times 10^4$
($\partial\Omega/\partial r<0$),
$\varpi_0=0.3$.}
\end{figure}

\begin{figure}[!thb]
\resizebox{\hsize}{!}{\includegraphics{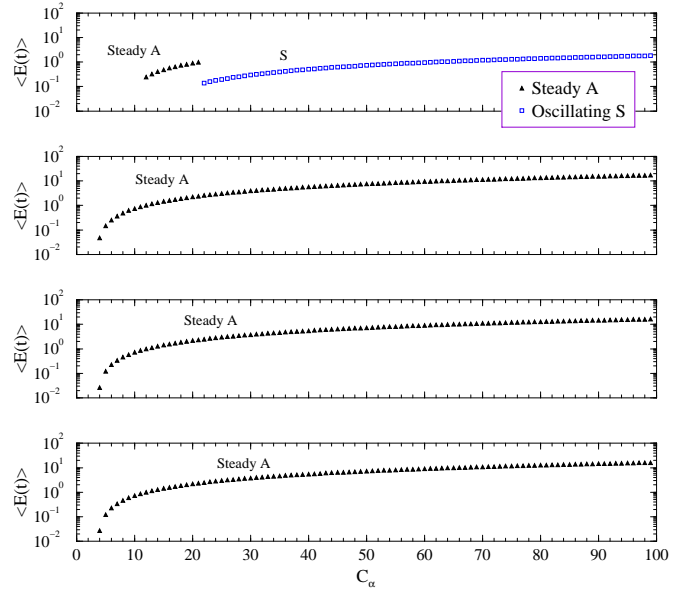}}
\caption{\label{keplerianminus}
Partial bifurcation  diagram, as Fig.\ \ref{none},
for the case of a large polar cut angle, $\theta_0=75^{\circ}$,
with negative generalized Keplerian rotation law,
$C_{\omega}=-10^4$ ($\partial\Omega/\partial r>0$),
$\varpi_0= 0.3$.}
\end{figure}

We present our results in the nonlinear regime in three stages.  First
we consider the transition from the full sphere to a spherical shell,
then from a full sphere to that with axial cones removed and finally we
study a full sphere with both an inner core and axial cones removed
simultaneously.  Our detailed results are presented in Figs.\
\ref{none} -- \ref{largeminus}, corresponding to five different values
of $\theta_0$. Each contains four panels for each of the values of
$r_0$ considered, the panels from top to bottom corresponding to $r_0$
values of $0.7, 0.5, 0.2$ and $0.0$ respectively.  We have concentrated
mostly on the more usual case of negative dynamo number
($C_\omega=-10^4$), but for comparison we also consider, for large
$\theta_0$, the case of a positive dynamo number ($C_\omega=+10^4$; see
Fig.\ \ref{largeplus}), as well as the case with generalized Keplerian
differential rotation, which is relevant for accretion discs (Figs.\
\ref{keplerianplus}--\ref{keplerianminus}).

\begin{figure}[!thb]
\resizebox{\hsize}{!}{\includegraphics{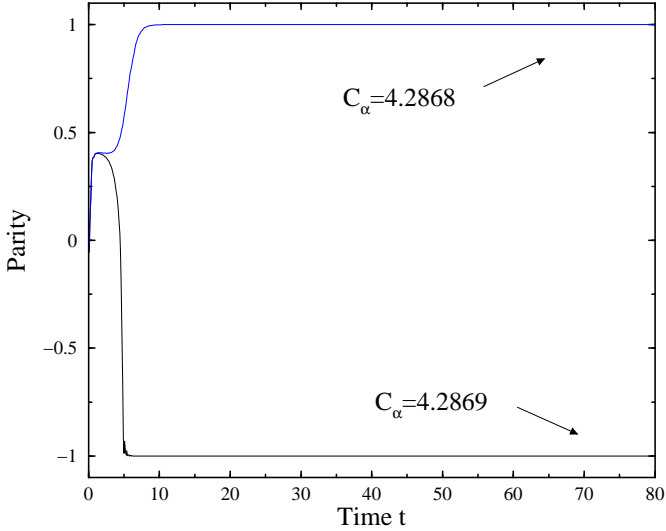}}
\caption{\label{paritytimes}
Parity versus time diagram for two runs with same initial conditions
except the value of $C_\alpha$. The other parameters used were
$C_\omega=-10^4$, $r_0=0.2$, $\theta_0=45^{\circ}$,
with a constant shear rotation profile.
}
\end{figure}

Before discussing our results in detail we emphasise that the sequence
of parity changes reported below are those that we found to within the
resolution of the $C_{\alpha}$ parameter employed here. It is therefore
important to bear in mind that unless there is a clear theoretical
reason for believing in the presence of a particular sequence of parity
changes in the nonlinear regimes, it is always possible to miss finer
details of such sequences for any given finite resolution in
$C_{\alpha}$. Our standard step in $C_\alpha$ is 0.5, but on occasion this
was very much reduced (see, e.g., Fig.\ \ref{paritytimes}).
In addition, although we have attempted to integrate over times long
enough to eliminate any transient effects,
nonetheless, it is not possible to always
rule out the presence of very long-lived transients.
With these qualifications we proceed to discuss our results
in more detail.

The results concerning the transition from a full sphere to spherical
shells are shown in Fig.\ \ref{none}. This shows that the transitions,
from the onset of dynamo action into the nonlinear regimes as
$C_{\alpha}$ increases, are the same for the full sphere and the two
thickest shells (i.e.\ $r_0=0.0,0.2,0.5$).  All possess the same initial
sequence of parity changes given by
\begin{equation}
A, OM, S, A.
\end{equation}
\noindent The only difference is that, as the shells become thinner ($r_0$
increases), the sequence is compressed to a narrower interval of
$C_{\alpha}$.  For the thinnest shell ($r_0 = 0.7$), however, the
initial sequence is
\begin{equation}
S, A, S, A.
\end{equation}

When a polar cone is removed ($\theta_0>0$), the initial sequence of
parity changes is the same for the full sphere and the two thickest
shells for $\theta_0 \le 45^\circ$. The behaviour for the thinnest
shell ($r_0=0.7$) and the largest polar cut angle ($\theta_0 = 75^{\circ}$)
differ.  We also observe that the removal of the polar cones causes more
changeable and varied behaviour, as does the removal of the inner cores,
providing that the first bifurcation is to an oscillating solution.

Figures \ref{very_small} to \ref{medium} show that there are similarities in the initial sequences of
parity changes as $\theta_0$ changes.
The configurations with $\theta_0= 4.5^{\circ}, 22.5^{\circ}$
and $45^{\circ}$ and $r_0=0.0, 0.2$ and $0.5$ all have the same initial
sequence, namely,
\begin{equation}
A,OM,S,A,
\end{equation}
whilst configurations with $r_0=0.7$ and these same
angles have the sequence
\begin{equation}
S, A, S, A.
\end{equation}
These two initial sequences agree with the corresponding ones in the
full sphere case ($\theta_0=0^{\circ}$). Differences in the bifurcation sequences
tend to appear at larger values of $C_\alpha$ where the dynamo becomes
more non--linear. Also at these higher values of $C_\alpha$,
more exotic behaviours such as intermittency and chaos can appear,
for example for $\theta_0=0^{\circ}$ and $4.5^{\circ}$ with $r_0=0.2$ and $0.5$,
$\theta_0=22.5^{\circ}$ with $r_0=0.2, 0.5$ and $0.7$, and
$\theta_0=45^
{\circ}$ for all values of $r_0$ considered here, whilst chaotic and
intermittent behaviour only occurs for $\theta_0=75^{\circ}$ when
$r_0=0.7$.

When $\theta_0=75^\circ$ (Fig.\ \ref{largeminus}) the behaviour is
totally different. The steady quadrupole mode ($S$) remains stable for a
very large range of values of $C_\alpha$. At some point an oscillatory
$S$ mode takes over.  For $r_0=0.5$ $OM$ modes appear and only for
$r_0=0.7$ are oscillatory modes observed for all values of $C_\alpha$
studied.

We also studied the effects of changing the sign of $C_\omega$ on our
results and this is shown in Fig.\ \ref{largeplus}. As can be seen, a
steady $A$ mode is first excited for all $r_0$. The general
behaviour is also less complex and more uniform, particularly
if we compare the panel corresponding to $r_0=0.7$ for both
signs of $C_\omega$.

The effects of changing the rotation law to the generalized Keplerian
rotation law are depicted in Fig.\ \ref{keplerianplus} and, as can be
seen, the steady $S$ mode and the subsequent oscillatory $S$ mode are
stable for all values of $C_\alpha$ considered. Only in the case
$r_0=0.7$ is new behaviour seen, in that there is an interval
($C_\alpha\approx15$) where only the trivial solution with zero
magnetic field exists. Such a gap in the bifurcation diagram has been
found earlier in connection with torus and accretion disc dynamos
(e.g.\ Brooke \& Moss 1995, Torkelsson \& Brandenburg 1994b, see also
Ruzmaikin {\em et al.} 1980  and Stepinski \& Levy 1991 
for linear results).  Immediately after this gap
an antisymmetric mode appears.  The existence of this gap does not
appear to be a peculiarity associated with the particular value of
$C_\omega$ we used, as we also found it when $C_\omega = 2.7 \times
10^4$, a value chosen to give the same value of the rotational shear at
the mean radius of the shell ($r=0.85$) as the constant shear model
with $C_\omega=-10^4$, shown in Fig.\ \ref{largeminus}. We show these
results in Fig.\ \ref{keplerianplus2}.  It is clear from comparing
Fig.\ \ref{keplerianplus} and Fig.\ \ref{keplerianplus2} that the
bifurcation sequence is sensitive to the value of $C_\omega$, but the
gap in which no field is excited remains.

When the sign of the generalized Keplerian rotational shear is reversed
(Fig.\ \ref{keplerianminus}), a steady $A$ mode is again first excited,
similar to the case of linear shear. Only for $r_0=0.7$ is the behaviour
slightly different in that there is a $S$ mode that appears for
$C_\alpha\approx22$.

In the above figures, the nature of the solutions (and thus the symbols
shown) are all determined only to within the numerical resolutions
chosen and for the length of of time during which solutions were
followed and thus in which transients could decay.  In particular, near
the bifurcation points (as in the case of transitions from pure parity A
to S (or vice versa)), we find as expected extremely long transients (at
times of more than hundreds of diffusion times).  Bearing all this in
mind, we do find some evidence which suggests that the conjecture by
Jennings \& Weiss (1991), according to which transitions from one pure
parity to the opposite are mediated by a mixed parity regime, may not
always hold.  An example of this type of transition is given in Fig.\
\ref{paritytimes}, where we have used an exceptionally fine resolution
in $C_\alpha$.  In this case we found after some hundreds of diffusion
times that with $C_\alpha=4.2868$ and $C_\alpha=4.2869$
the parity became $P=+0.9999633$ and
$P=-0.9999967$ respectively. 
A conclusive resolution of this question requires a
more substantial study, and we intend to return to this point in a
future publication.

Our results also indicate that overall the behaviour appears to be more
complicated in shells than in the full sphere dynamo models, a feature
also observed in Covas {\em et al.} 1998a.

Also, the nonlinear solutions in configurations which correspond most
closely to a thick torus, i.e.\ $\theta_{0} = 22.5^{\circ}$,
$r_{0}$ = 0.2, 0.5 and also $\theta = 45^{\circ}$, $r_{0} = 0.5$,
display chaotic behaviour at the lowest values of $C_{\alpha}$.  This
corresponds with the findings of Brooke \& Moss (1994, 1995), that
chaotic behaviour was excited at modestly supercritical dynamo numbers
in tori, whereas chaotic behaviour only occurred at highly supercritical
dynamo numbers in disc and shell configurations.

\begin{figure}[!thb]
\resizebox{\hsize}{!}{\includegraphics{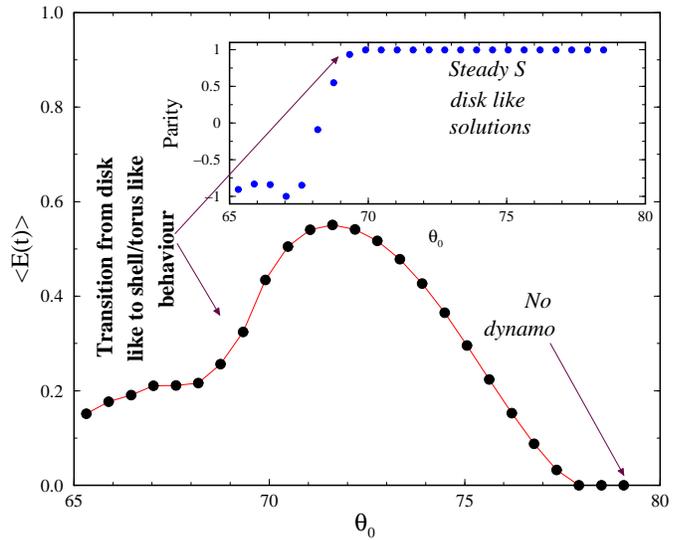}}
\caption{\label{transition.sphere-shell}
Transition from sphere/shell to disc-like behaviour
for a constant shear profile
as a function of $\theta_0$
for $C_{\alpha} =10$, $r_0=0.5$, $C_{\omega}=-10^4$.
The inset gives the corresponding
behaviour of the parity.}
\end{figure}

\begin{figure}[!thb]
\resizebox{\hsize}{!}{\includegraphics{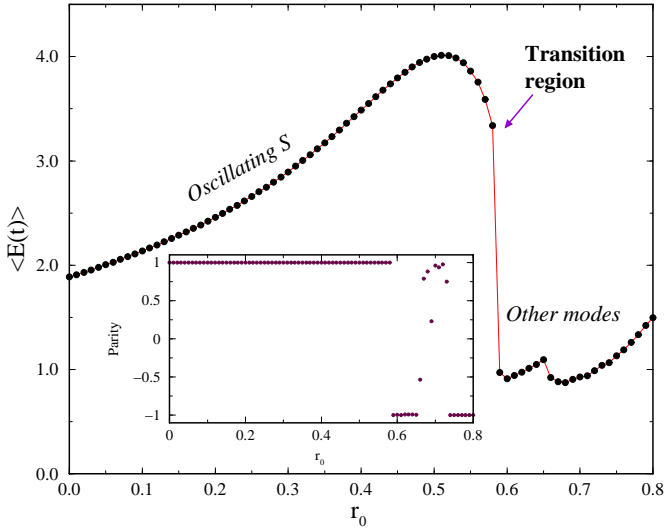}}
\caption{\label{transition.planar}
Transition from disc-like to planar like behaviour for a constant
shear profile as a function of $r_0$ for $C_{\alpha} =60$, $\theta_0
=75^\circ$, $C_{\omega}=-10^4$.  The inset gives the corresponding
behaviour of the parity.
}
\end{figure}

To obtain an overview of the behaviour of the two parameter family of
models considered here, we recall that Table \ref{onsetenergy} suggests
the presence of three different regions with distinct onset behaviours
in the parameter space $(r_0, \theta_0)$. In the constant shear rotation
law case two of these regions possess oscillatory dipolar and
quadrupolar onset symmetries respectively and in the third the most
readily excited field is steady and quadrupolar.  These regions could
also be interpreted from a different point of view by considering the
behaviour observed in previous dynamo models. Viewed in this way, there
are at least two transitions in the $(r_0, \theta_0)$ parameter space.
The first corresponds to a transition from a region which is consistent
with Roberts' (1972) results (characterised by an $A$ onset mode for
$r_0 < 0.7$ and by an $S$ onset mode with $r_0 \ge 0.7$) to one that
does not (i.e.\ one with antisymmetric onset modes for all $r_0$). The
second transition is from a disc--like regime in which steady modes are
commonly found to one in which they are absent.  In order to elucidate
this second transition we have made two studies of the parameter space
at a finer resolution, by fixing $r_0$ and varying $\theta_0$ and vice
versa.  These results are depicted in Figs.\
\ref{transition.sphere-shell} and \ref{transition.planar} respectively
at fixed values of $C_\alpha$. (We note that changing the value of
$C_\alpha$ does not qualitatively change our conclusions.) Those results
indicate the presence of two sharp transitions in mean energy and/or
parity at around $\theta_0 \sim 70^{\circ}$ for $r_0 \sim 0.5$ and
$\theta_0 \sim 75^{\circ}$ for $r_0 \sim 0.6$.  These features are also
corroborated by the time series of energy and parity.

Finally, we have also
examined the behaviour of
our solutions as a function of the aspect ratio, which
we defined as
\begin{equation}
A= \frac{2(1-r_0)} {(1+r_0) (\pi-2 \theta_0)}.
\end{equation}

With this definition, the smallest aspect ratio corresponds to the
configuration with $r_0=0.7$ and $\theta_0=0$ ($A=0.11$), and the
largest to that with $r_0=0$ and $\theta_0=75^\circ$ ($A=3.82$). We find
that at large values of $C_\alpha$, for example $C_\alpha > 30$, the
averaged energy $\langle E(t)\rangle$ per unit volume of the dynamo
region is largest for the configuration with the smallest aspect ratio,
and vice versa.
In addition, for these values of $C_\alpha$, we find that the average
energy per unit volume decreases as $\theta_0$ increases (i.e.\ the cuts
get larger) and as $r_0$ decreases (i.e.\ the shells get thicker).
\section{Conclusions}
We have made a comparative study of the dynamical behaviour of a two
parameter family of axisymmetric mean-field dynamo models, ranging from
a full sphere to shells to torus- and disc-like configurations, within a
unified numerical framework. We have found overall agreement with
previous studies of axisymmetric mean-field dynamos for spheres, shells,
tori and discs.  More precisely we find the following:
\begin{enumerate}
\item
Our results show that changes in both topology and geometry affect the
dynamical properties of our models, although these two factors are
inevitably interlinked.  Specifically, we have three topologically
distinct configurations, sphere-like, shell-like and torus-like. Each of
these topological configurations can undergo a continuous sequence of
geometric changes, from flat objects where the vertical extent is much
less than the horizontal, to objects where the vertical and horizontal
extents are similar. Our principal findings are
that torus-like topologies seem to produce more complex behaviour and
that, within a given topological configuration, modest changes in
geometry can give rise to different bifurcation sequences. These
findings may have considerable astrophysical significance, in that the
actual shapes of galaxies and accretion discs often do not conform to
the simple geometric shapes used in many studies of mean-field dynamos.
Also, since our study concerns the geometry and topology of the
dynamo-active region, restrictions of the dynamo-active region in
stellar convection zones may have important consequences for the
behaviour of the dynamo and these should be taken into consideration
when comparing the predictions of dynamo models with observed magnetic
behaviour in astrophysical objects.
\item
We confirm that the full sphere and shell results of Roberts (1972),
concerning the symmetry of the onset modes,
are robust with respect to changes in the outer boundary condition,
and we extend them to the
configurations with very small to medium polar cut angles.  Steady modes are
found for the largest polar cut angles, agreeing with results from
oblate spheroidal geometry. These modes are robust under a change in the
rotation law, as found by Walker \& Barenghi (1994), except in the case
where the configuration resembles a thin torus rather than a disc. In
this case an oscillating dipolar solution occurs at onset for a constant
shear rotation law, but a steady quadrupolar solution is found with a
generalized Keplerian law. In the latter case the steady solution
disappears at higher values of $C_\alpha$ and there is a gap before an
oscillating antisymmetric solution appears.  Such a gap was previously
observed with a generalized Keplerian rotational profile in both disc
(Torkelsson \& Brandenburg 1994b) and torus geometry (Brooke \& Moss
1995). Stepinski \& Levy (1991) also found a similar gap in linear theory.
It should be noted that the presence of a gap does depend on the
nature of the nonlinearity: it is absent if magnetic buoyancy is the
dominant nonlinearity (Torkelsson \& Brandenburg 1994a).
Our results here indicate the steady solution is also fragile
with respect to changes in the rotation law.
\item
For the largest polar cut angle, we find evidence for a different mode
of behaviour in the thinnest shell ($r_0=0.7$) from that found in the
thicker shells and full sphere, namely a more complicated behaviour with
denser bifurcation sequences, including chaotic and intermittent
solutions.  Apart from the thinnest shells ($r_0=0.7$) the dynamics is
qualitatively the same for each sign of the shear. This is consistent
with the observation that the thinnest shell approximates a setting in
which curvature plays a smaller role.
\item
We note a consistent finding in our results, that dynamo-active regions
corresponding topologically and geometrically to a torus exhibit chaotic
behaviour at lower dynamo numbers compared to spherical, shell-like or
disc-like configurations. This concurs with the results of Brooke
\& Moss (1994,1995), who found transition to chaotic behaviour for a
dynamo in a toroidal volume at dynamo numbers that were approximately
three times supercritical. Given that the solar (and possibly some
stellar) magnetic cycle exhibits irregularity on a variety of time
scales, and that many authors have speculated that such behaviour is a
manifestation of deterministic chaos, this result may have some
astrophysical significance. It therefore seems worthwhile to investigate
such dynamos, with a variety of rotational profiles, possibly corresponding to
different types of stars, and using different forms of nonlinearity to
saturate the field. Since it is the restriction of the dynamo active
volume to a toroidal configuration that appears to be important, these
dynamos could be considered as embedded in convective shells
representing the outer regions of stars.
\end{enumerate}
\begin{acknowledgements}
EC is supported by grant BD / 5708 / 95 -- Program PRAXIS XXI, from
JNICT -- Portugal.  RT benefited from PPARC UK Grant No. L39094.
Support from the EC Human Capital and Mobility (Networks) grant ``Late
type stars: activity, magnetism, turbulence'' No. ERBCHRXCT940483 is
also acknowledged.
\end{acknowledgements}


\begin{thebibliography}{}
\bibitem{betal}
Brandenburg, A., 1999, in: {\it Theory of Black Hole Accretion Discs},
eds. M. A. Abramowicz, G. Bj\"ornsson \& J. E. Pringle,
Cambridge University Press

\bibitem{brandenburg89a}
Brandenburg, A., Krause, F., Meinel, R., Moss, D.\ \& Tuominen, I., 1989a,
A\&A {\bf 213}, 411

\bibitem{btm}
Brandenburg, A., Moss, D.\ \& Tuominen, I., 1989b, Geophys. Astrophys. Fluid
Dyn. {\bf 40}, 129

\bibitem{brandenburg95}
Brandenburg, A., Nordlund, \AA., Stein, R. F., Torkelsson, U., 1995,
ApJ {\bf 446}, 741

\bibitem{brandenburg96}
Brandenburg, A., Jennings, R. L., Nordlund, \AA., Rieutord, M.,
Stein, R. F., Tuominen, I., 1996, JFM {\bf 306}, 325

\bibitem{brooke94}
Brooke J.\ M.\ \& Moss D., 1994, MNRAS {\bf 266}, 733

\bibitem{b95}
Brooke, J. M. \& Moss, D., 1995, A\&A {\bf 303}, 307

\bibitem{covas98a} Covas,  E., Tavakol, R., Tworkowski, A. \&
Brandenburg, A., 1998a, A\&A {\bf 329}, 350

\bibitem{covasetal98} Covas, E., Tavakol, R., Ashwin, P., Tworkowski, A.\
\& Brooke, J.\ M., 1998b, submitted to PRL

\bibitem{gilman} Gilman, P.A. \& Miller, J., 1981, ApJ Suppl.
{\bf 46}, 211

\bibitem{hawley}
Hawley, J.F., Gammie, C.F., \& Balbus, S.A., 1996, ApJ {\bf 464}, 690

\bibitem{jennings1991a}
Jennings, R.L., 1991, Geophys. Astrophys. Fluid Dyn. {\bf 57}, 147

\bibitem{jennings1991b}
Jennings, R.L., Weiss, N.O., 1991, MNRAS {\bf 252}, 249

\bibitem{jennings1991c}
Jennings, R.L., Brandenburg, A., Moss, D. \& Tuominen, I.,
1990, A\&A {\bf 230}, 463

\bibitem{jones}
Jones, C.A., Weiss, N.O., Cattaneo, F., 1985, {\em Physica 14D}, 161

\bibitem{kitchatinov} Kitchatinov, L.L., 1987, Geophys.\ Astrophys.\
Fluid Dyn. {\bf 38}, 273

\bibitem{krause} Krause, F. \& R\"adler, K.-H., 1980, {\it Mean-Field
Magnetohydrodynamics and Dynamo Theory}, Pergamon Press, Oxford

\bibitem{nordlund}
Nordlund, \AA., Brandenburg, A., Jennings, R. L., Rieutord, M.,
Ruokolainen, J., Stein, R.F., Tuominen, I., 1992, ApJ {\bf 392}, 647

\bibitem{radler86}
R\"adler, K.-H. 1986, Plasma Physics, ESA SP-251, 569

\bibitem{roberts72a}
Roberts, P. H., 1972, Phil. Trans. Roy. Soc. {\bf A272}, 663

\bibitem{roberts72b}
Roberts, P.H., Stix, M., 1972, A\&A {\bf 18}, 453

\bibitem{ruzmaikin}
Ruzmaikin A.\ A., Sokoloff D.\ D., Turchaninov V., 1980, Sov.\
Astron.\ {\bf 24}, 182

\bibitem{soward1} Soward, A.,  1992a, Geophys. Astrophys. Fluid Dyn.,
{\bf 64}, 163

\bibitem{soward2} Soward, A.,  1992b, Geophys. Astrophys. Fluid Dyn.,
{\bf 64}, 201

\bibitem{steenbeck}
Steenbeck, M., Krause, F., 1969, Astron. Nachr. {\bf 291}, 49

\bibitem{stepinski}
Stepinski T.F., Levy E.H., 1991, ApJ {\bf 379}, 343

\bibitem{stix} Stix, M., 1975, A\&A {\bf 42}, 85

\bibitem{tavakoletal} Tavakol, R.K., Tworkowski, A.S., Brandenburg,
A., Moss, D. \& Tuominen, I., 1995, A\&A {\bf 296}, 269

\bibitem{tobias95}
Tobias, S.M., Weiss, N.O., Kirk, V., 1995, MNRAS {\bf 273}, 1150

\bibitem{tb94a}
Torkelsson, U. \& Brandenburg, A., 1994a, A\&A {\bf 283}, 677

\bibitem{tb94b}
Torkelsson, U. \& Brandenburg, A., 1994b, A\&A {\bf 292}, 341

\bibitem{tb95}
Torkelsson, U. \& Brandenburg, A., 1995, Chaos, Solitons \& Fractals
{\bf 5}, 1975

\bibitem{tworkowskietal} Tworkowski, A., Tavakol, R., Brandenburg, A.,
Brooke, J.\ M., Moss, D. \& Tuominen, I., 1998, MNRAS {\bf 296}, 287.

\bibitem{walker}
Walker, M.R., Barenghi, C.F., 1994, Geophys. Astrophys. Fluid Dyn.,
{\bf 76}, 265

\bibitem{weiss84}
Weiss, N.O., Cattaneo, F., Jones, C.A., 1984, Geophys. Astrophys. Fluid
Dyn. {\bf 30}, 305

\bibitem{zeldovich} Zeldovich, Ya.B., Ruzmaikin, A.A. \& Sokoloff,
D.D., 1983: {\it Magnetic Fields in Astrophysics}, Gordon and Breach,
New York.
\end{thebibliography}
\end{document}